\documentclass[fleqn, journal=jacsat,manuscript=article]{achemso}
\setkeys{acs}{articletitle = true}
\setkeys{acs}{etalmode=truncate,maxauthors = 10}

\usepackage[version=3]{mhchem} 
\usepackage{booktabs}
\usepackage{tabularx}
\usepackage[colorinlistoftodos]{todonotes}
\usepackage{color}
\usepackage{setspace}
\usepackage{color,soul}
\usepackage{xcolor}
\usepackage{amsmath}
\usepackage{mathtools}
\usepackage{amsmath}
\usepackage{soul}

\SectionNumbersOn

\author{Somnath Biswas}
\affiliation[Princeton University]
{Department of Chemistry, Princeton University, Washington Road, Princeton, New Jersey 08544, United States}

\author{Fatimah Alowa}
\affiliation[Boston University]
{Division of Materials Science and Engineering, Boston University, Boston, Massachusetts 02215, United States}

\author{Ruyan Zhao}
\affiliation[University of Toronto]
{Department of Chemistry, University of Toronto, 80 St. George Street, Toronto, Ontario M5S 3H6, Canada}

\author{Marios Zacharias}
\affiliation{Univ Rennes, INSA Rennes, CNRS, Institut FOTON - UMR 6082, F-35000 Rennes, France}

\author{Sahar  Sharifzadeh}
\affiliation[Boston University]
{Division of Materials Science and Engineering, Boston University, Boston, Massachusetts 02215, United States}
\affiliation[Boston University]
{Department of Electrical and Computer Engineering, Boston University, Boston, Massachusetts 02215, United States}
\author{David F. Coker}
\affiliation[Boston University]
{Department of Chemistry, Boston University, Boston, Massachusetts 02215, United States}
\author{Dwight S. Seferos}
\affiliation[University of Toronto]
{Department of Chemistry, University of Toronto, 80 St. George Street, Toronto, Ontario M5S 3H6, Canada}
\author{Gregory D. Scholes*}
\affiliation[Princeton University]
{Department of Chemistry, Princeton University, Washington Road, Princeton, New Jersey 08544, United States}
\setkeys{acs}{email = false}

\title[An \textsf{achemso} demo]
 {Exciton-phonon Coupling Controls Exciton-polaron Formation and Hot Carrier Relaxation in Rigid Dion-Jacobson Type Two-Dimensional Perovskites}

\begin{document}
\maketitle

\newpage
\begin{abstract}

The efficiency of two-dimensional Dion-Jacobson-type materials relies on the complex interplay between electronic and lattice dynamics; however, questions remain about the functional role of exciton-phonon interactions. This study establishes the robust polaronic nature of the excitons in these materials at room temperature by combining ultrafast spectroscopy and electronic structural calculations. We show that polaronic distortion is associated with low-frequency (30-60 cm$^{-1}$) lead iodide octahedral lattice motions. More importantly, we discover how targeted ligand modification of this two-dimensional perovskite structure controls exciton-phonon coupling, exciton-polaron population, and carrier cooling. At high excitation density, stronger exciton-phonon coupling increases the hot carrier lifetime, forming a hot-phonon bottleneck. Our study provides detailed insight into the exciton-phonon coupling and its role in carrier cooling in two-dimensional perovskites relevant for developing emerging hybrid semiconductor materials with tailored properties.

\end{abstract}


\newpage

\noindent {\large{\textbf{Introduction}}}

Two-dimensional (2D) hybrid organic-inorganic lead-halide perovskites (HOIP) have attracted considerable attention because of the strongly confined nature of their excitonic states and the reduced dielectric screening resulting from their 2D layered structure enabling applications in light-emitting devices, photodetectors, photovoltaics, and quantum emitters.\cite{blancon2017extremely,tsai2016high,zhao2020large,wang2021low} 
Insight into the coupling between lattice vibrations and excitons
may reveal the polaronic nature of the excitations, which can improve photovoltaic device performance.\cite{zhu2016screening,miyata2017large,niesner2016persistent,buizza2021polarons}

The growing interest in incorporating these materials in hot-carrier solar cells stems from the profound exciton-phonon interaction that controls the hot-carrier cooling kinetics.\cite{yin2019tuning,yin2021manipulation} 
Therefore, systematic control of exciton-phonon coupling and understanding its influence on carrier cooling is critical to tune the performance of these materials.

While it is evident that the efficiency of 2D perovskites depends strongly on the complex interplay between the electronic and lattice/structural dynamics, a clear understanding of exciton-phonon interaction and their role in device performance is still an open question.\cite{mayers2018lattice,silva_jpcl_2020,Thouin2019} 
In particular, this understanding is largely unexplored in the case of Dion-Jacobson (DJ)-type 2D materials where organic ligands serve as the bifunctional di-cations and provide a robust and rigid perovskite framework.\cite{li2019two,zhao2021rigid}
%
This 2D architecture is more stable than the Ruddlesden-Popper(RP) type perovskites, another subset of 2D-HOIPs incorporating monofunctional layers of organic mono-cationic ligands which result in reduced relative material stability. \cite{stoumpos2016ruddlesden}
Recent, femtosecond electron diffraction measurements reveal distinctly different carrier-lattice interactions between DJ- and RP-type 2D perovskites, demonstrating the importance of understanding exciton-phonon interaction in these systems. \cite{zhang2023ultrafast} 
Due to strong carrier-lattice interaction, photoexcitation in these 2D perovskite structures is strongly coupled to the lattice distortion along certain phonon coordinates to form exciton-polaron states.\cite{silva_jpcl_2020, Thouin2019}

Efforts to understand the excitonic physics in RP-type perovskites reveal the polaronic character of the excitons in these materials where the motions of the inorganic octahedra (\ce{PbI6}) provide a complex energetic landscape for the charges to induce polaron formation.\cite{silva_jpcl_2020, Thouin2019,Fu2021}
However, an analogous understanding of excitonic nature and exciton-phonon coupling in more stable and robust DJ-type 2D-HOIPs has yet to be investigated. 
More importantly, controlling the exciton-phonon interaction through targeted ligand engineering in this class of materials to tune the carrier cooling can significantly impact the applicability of these materials in hot carrier solar cell applications, for example.\cite{zhao2021rigid}

In the study described here, we employ time-resolved spectroscopy to provide evidence of exciton-polaron formation at room temperature in rigid DJ-type perovskites by observing molecular-like phonon wave-packet motion. 
The coherent phonons are observed to modulate the excitonic resonance, indicating the strong exciton-phonon coupling regime operating in this class of materials.
These conclusions are supported by electronic structure calculations using the special displacement method~\cite{ZG_2016,ZG_2020} where we show that phonon modes between 30-60 cm$^{-1}$ renormalize the excitonic transition energy.
Our study suggests that polaronic distortion of the exciton occurs along these specific phonon coordinates.
Additionally, we show that the exciton-phonon coupling can be tuned through ligand modification which controls the hot carrier lifetime, exciton-polaron population, and phonon bottleneck formation.
We find that an increase in the exciton-phonon coupling strength increases the hot carrier lifetime at higher excitation density ($>10 ^{18} cm^{-3}$), suggesting the formation of a hot phonon bottleneck. 

\noindent {\large{\textbf{Exciton-polaron formation through wavepacket dynamics}}}

\begin{figure*}[ht]
\includegraphics[width=6in]
{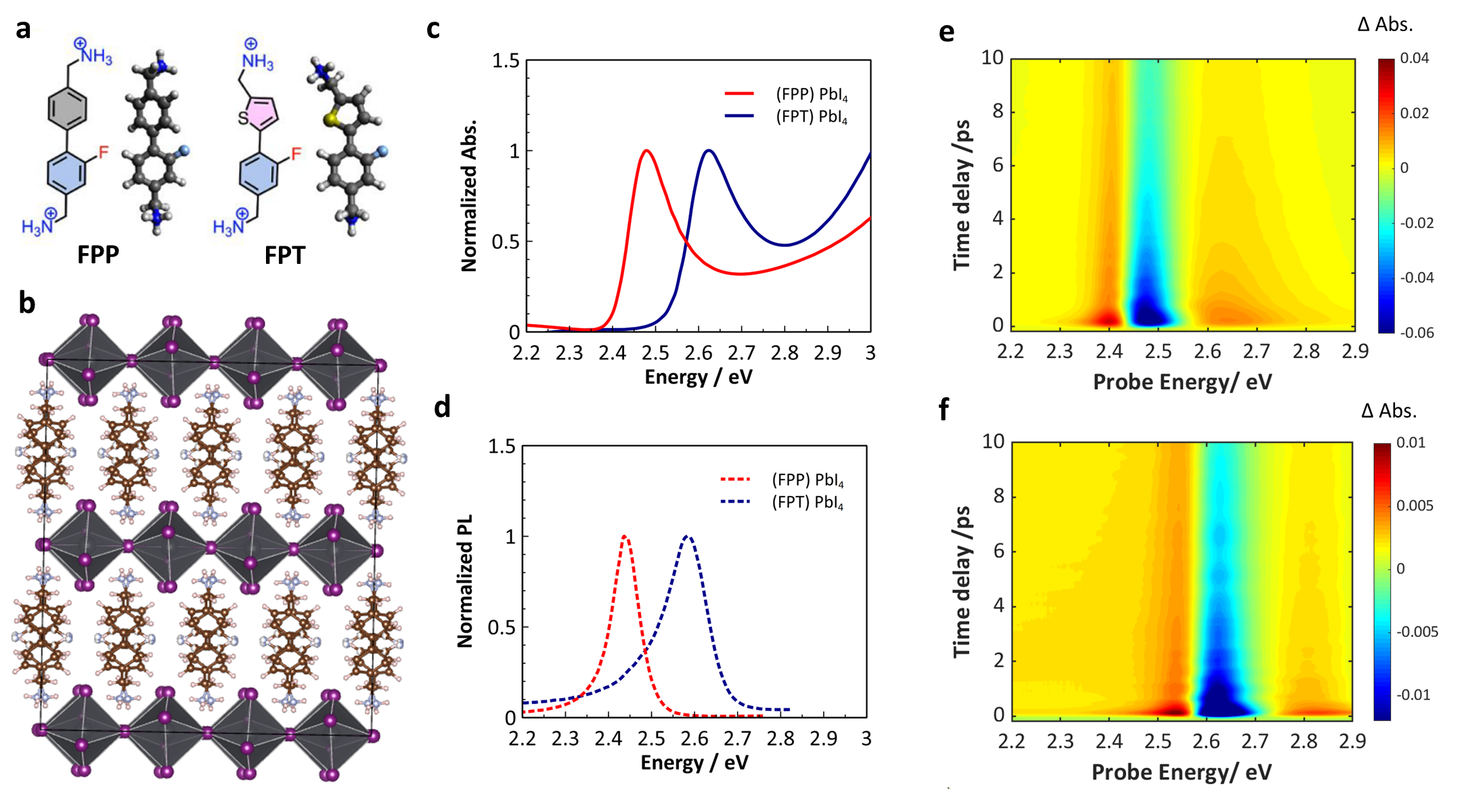}
\centering
 \caption {(a) Structure of bifunctional organic ligands used to synthesize the DJ type perovskites  (b) DJ-type perovskites, showing the layered structure (c) Absorption spectra of both (FPP)\ce{PbI4} and (FPT)\ce{PbI4}, showing the excitonic resonance at 2.47 eV and 2.63 eV, respectively. (d) Emission spectra of both (FPP)\ce{PbI4} and (FPT)\ce{PbI4}. (e,f) Transient absorption signal of both (FPP)\ce{PbI4} and (FPT)\ce{PbI4} following photoexcitation above the band gap at 3.1 eV obtained at 20.8 and 12.5 $\mu$J/cm$^{-2}$ pump fluence, respectively.}
  \label{Structure_Abs_Em_TA}
\end{figure*}

Figure \ref{Structure_Abs_Em_TA}a shows the structure of the organic ligands used to synthesize the DJ-type 2D perovskites: (1) (FPP)\ce{PbI4} (FPP : 2-fluoro-[1,1'-biphenyl]-4,4'-diyl)-dimethanaminium), and (2) (FPT)\ce{PbI4} (FPT : 5-(4-(ammoniomethyl)-2-fluorophenyl)thiophen-2-yl)methanaminium). 
 Structure of (FPP)\ce{PbI4} is displayed in Figure \ref{Structure_Abs_Em_TA}b. 
Details of the material synthesis have been reported before and a brief description is provided in section 1 of the supplementary materials.\cite{zhao2021rigid}
We note clear excitonic resonances in these different materials at 2.47 eV and 2.62 eV in the absorption spectra (Figure \ref{Structure_Abs_Em_TA}c); and emission maxima at 2.44 eV and 2.59 eV for (FPP)\ce{PbI4} and (FPT)\ce{PbI4}, respectively (Figure \ref{Structure_Abs_Em_TA}d). 
Blue-shifting the excitonic transition energy by $\sim$0.15 eV indicates the presence of significantly distorted inorganic octahedra in the case of (FPT)\ce{PbI4}, consistent with our electronic structure calculations (Figure S6 of supplementary materials) and previous reports for Pb-I based 2D and 3D perovskites.\cite{Smith2017,filip2014steric}
The octahedral distortion reduces the overlap between the Pb and I orbitals, resulting in lower band dispersion and a larger band gap.\cite{chakraborty2023rational,wang2020structural} 
Additionally, the observation of a broad PL lineshape indicates the presence of larger exciton-phonon coupling in (FPT)\ce{PbI4} compared to (FPP)\ce{PbI4}.\cite{ni2017real,wright2016electron} 
Therefore, (FPT)\ce{PbI4} and (FPP)\ce{PbI4} are suitable model systems to develop a comprehensive understanding of the complex interplay between excitons and phonons and their effects on the exciton-polaron formation and hot carrier relaxation in 2D DJ type perovskites systems. 

Figure \ref{Structure_Abs_Em_TA}e shows the transient absorption (TA) data for (FPP)\ce{PbI4} as a function of probe energy and time delay following photoexcitation with a 410 nm pump (3.1 eV).
Experimental details of the transient absorption measurements are provided in section 2 of the supplementary materials.
In the TA data for (FPP)\ce{PbI4} (Fig.~\ref{Structure_Abs_Em_TA}e)  we observe a negative feature  at 2.47eV and positive features both at lower and higher energies relative to the negative feature, consistent with previous reports.\cite{yin2021manipulation, Fu2021}
The negative feature in Figure \ref{Structure_Abs_Em_TA}e is right at the energy of the ground state absorption of (FPP)\ce{PbI4}, and is therefore associated with the ground state bleach of the exciton.\cite{yang2016observation,manser2014band} 
We assign the positive feature below the band gap region (2.4 eV) to ultrafast band gap renormalization, and the fast decay of this feature is associated with hot carrier relaxation as described below.
Carriers in the conduction band induce a many-body effect as a result of carrier-carrier exchange and correlations, producing a reduction of the band gap. 
As a result of this new band gap, we observe this photo-induced absorption feature (2.4 eV) which is red-shifted relative to the ground state excitonic absorption (2.47 eV).
As the carriers thermalize, and electrons occupy these sub-bandgap states, this photoinduced absorption feature recovers during carrier cooling.

To confirm this assignment we have performed an excitation wavelength dependence study where we note that the below-band gap positive absorption feature vanishes for near-resonant excitation (See section S3, figure S3).
Therefore, hot carriers with excess energy are responsible for this feature.
Our assignment is in agreement with several other reports in the literature on perovskites and other 2D materials.\cite{yang2016observation,ghosh2018reflectivity,wood2020evidence}
The above band-gap positive feature is consistent with photo-induced broadening of the absorption and has been assigned to the photo-induced change in the refractive index or to the photo-induced stark effect.\cite{price2015hot,tran2020observation}
We note similar TA spectral features for the case of (FPT)\ce{PbI2} sample (Figure \ref{Structure_Abs_Em_TA}f) where the exciton bleach appears at 2.63 eV, consistent with the blue-shifted ground state excitonic absorption for this material (Figure \ref{Structure_Abs_Em_TA}c) compared to (FPP)\ce{PbI2}.

\begin{figure*}[ht]
\includegraphics[width=6in]
{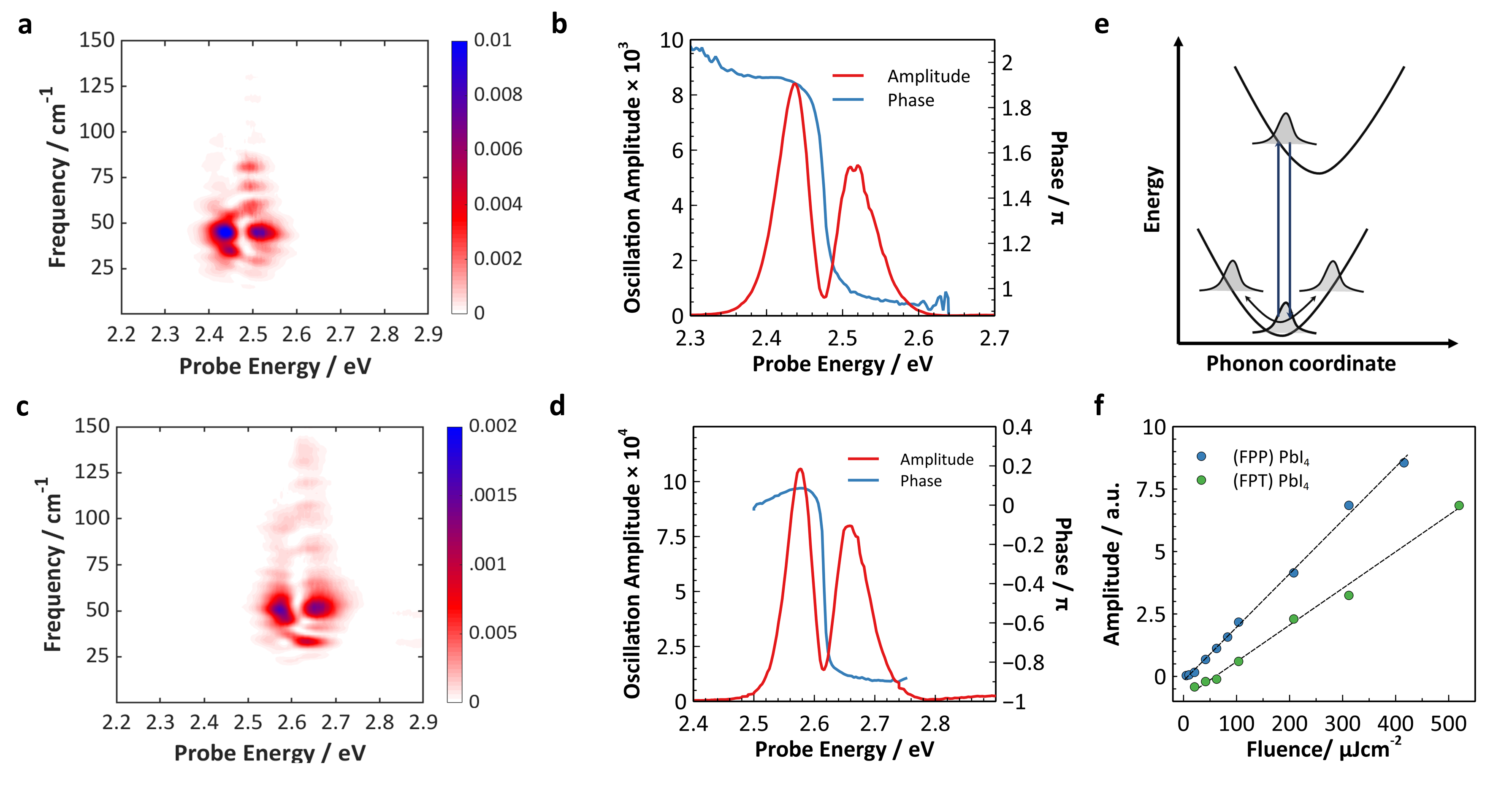}
\centering
 \caption {(a,c) Fourier transformed amplitude of the coherent phonon modulation as a function of time and probe energy for (FPP)\ce{PbI2} and (FPT)\ce{PbI2}, respectively. (b,d) Phase and amplitude of the phonon modulation at 50 cm$^{-1}$ as a function of probe energy for (FPP)\ce{PbI2} and (FPT)\ce{PbI2}, respectively, showing the presence of a node and phase jump. (e) Schematic of ISRS mechanism for coherent phonon generation (f) The amplitude of coherent phonon feature scaled as in (b,d) as a function of pump fluence, showing linear dependence.
 }
  \label{Exciton_Phase_Amp}
\end{figure*}

A closer look at the TA data reveals a coherent modulation of the transient signals both above and below the exciton bleach.
In order to obtain the frequency of the modulation we have subtracted the population decay and Fourier transformed the signals to obtain the results in Figure \ref{Exciton_Phase_Amp}a and c.
We note the presence of strong coherent oscillation centered around 45cm$^{-1}$ in the probe energy range 2.4 eV to 2.6 eV for the case of (FPP)\ce{PbI2} (Figure \ref{Exciton_Phase_Amp}a).
The phase and amplitude of the oscillation near 45cm$^{-1}$ as a function of probe energy are shown in Figure \ref{Exciton_Phase_Amp}b.
We observe a minimum (node) in the amplitude spectrum and a sharp $\pi$ phase jump at 2.47 eV, which is very close to the exciton resonance for (FPP)\ce{PbI2}.

The presence of such node and phase jump features can be explained in terms of the formalism developed for wavepackets and are typically observed in the condensed phase, {\em e.g.} in photosynthetic proteins, and organic conjugated systems.\cite{jumper2016broad,yan1990femtosecond,cina2016ultrafast}
When the ground and excited state potential energy surfaces are displaced relative to one another along the vibrational coordinates impulsive excitation using an ultrashort pulse generates vibrational wavepackets that oscillate around the global minima of the potential energy surface.
A key feature of the wavepacket oscillation is that the phase of the wavepacket undergoes a $\pi$ phase shift at a probe energy that corresponds to the global minima of the potential energy surface.
Interestingly, here we note a phase shift of $\pi$ near the exciton resonance (Figure \ref{Exciton_Phase_Amp}b). 
In fact, the extracted coherent oscillation below and above the exciton resonance at a particular probe energy shows almost a $\pi$ phase shift (see section S4 of supplementary materials) with cosine-like behavior.
Therefore, our analysis confirms the presence of molecular-like coherent phonon wavepacket motion in DJ-type 2D perovskites.
We also observed the presence of similar coherent phonons, (Figure \ref{Exciton_Phase_Amp}c), wavepacket node, and phase jump (Figure \ref{Exciton_Phase_Amp}d) in case of (FPT)\ce{PbI4}.
The position of the wavepacket node and the phase jump has blue shifted to around 2.63 eV, consistent with the blue-shifted excitonic resonance observed in (FPT)\ce{PbI4} (Figure \ref{Structure_Abs_Em_TA}c).
Similar wavepacket nodes have been observed before in 2D RP-type perovskite materials at very low temperatures.\cite{Thouin2019}
However, due to the strong vibrational dephasing, this wavepacket node was difficult to observe at room temperature in the 2D RP-type perovskites.
Here we show the first evidence of a vibrational wavepacket node at room temperature in DJ-type 2D perovskites materials.
The rigid platform provided by the DJ-type structure reduces the phonon dephasing rate, enabling the observation of coherent phonons at room temperature.\cite{zhao2021rigid}

The observed cosine-like behavior (see Figure S4 of supplementary materials) of the coherent phonons along with their symmetry (both A and B type) from electronic structure calculations (See Figures S9 and S10) suggest that impulsive stimulated Raman scattering (ISRS) is the primary mechanism of coherent phonon generation.
Additionally, a clear linear relationship between the amplitude of the coherent phonons and pump fluence in both samples (Figure \ref{Exciton_Phase_Amp}f) further confirms the ISRS mechanism.
The impulsive excitation of coherent phonons creates modulation via two different processes: (1) amplitude modulation (non-Condon effects)\cite{kambhampati2000solvent}, and (2) frequency modulation (exciton-phonon coupling)\cite{pollard1992theory,gambetta2006real}.
The oscillation as a function of probe energy is in phase for amplitude modulation, however, is out-of-phase for frequency modulation (see details in section S4 of supplementary materials).
In our experimental data, we observe a clear out-of-phase relationship ($\pi$ phase shift) below and above the excitonic resonance.
This analysis confirms that coherent phonons modulate the excitonic transition energy as a result of exciton-phonon coupling that appears as a frequency modulation in our transient absorption measurements, indicating the formation of exciton-polarons in these materials.


\begin{figure*}[ht]
\includegraphics[width=6in]
{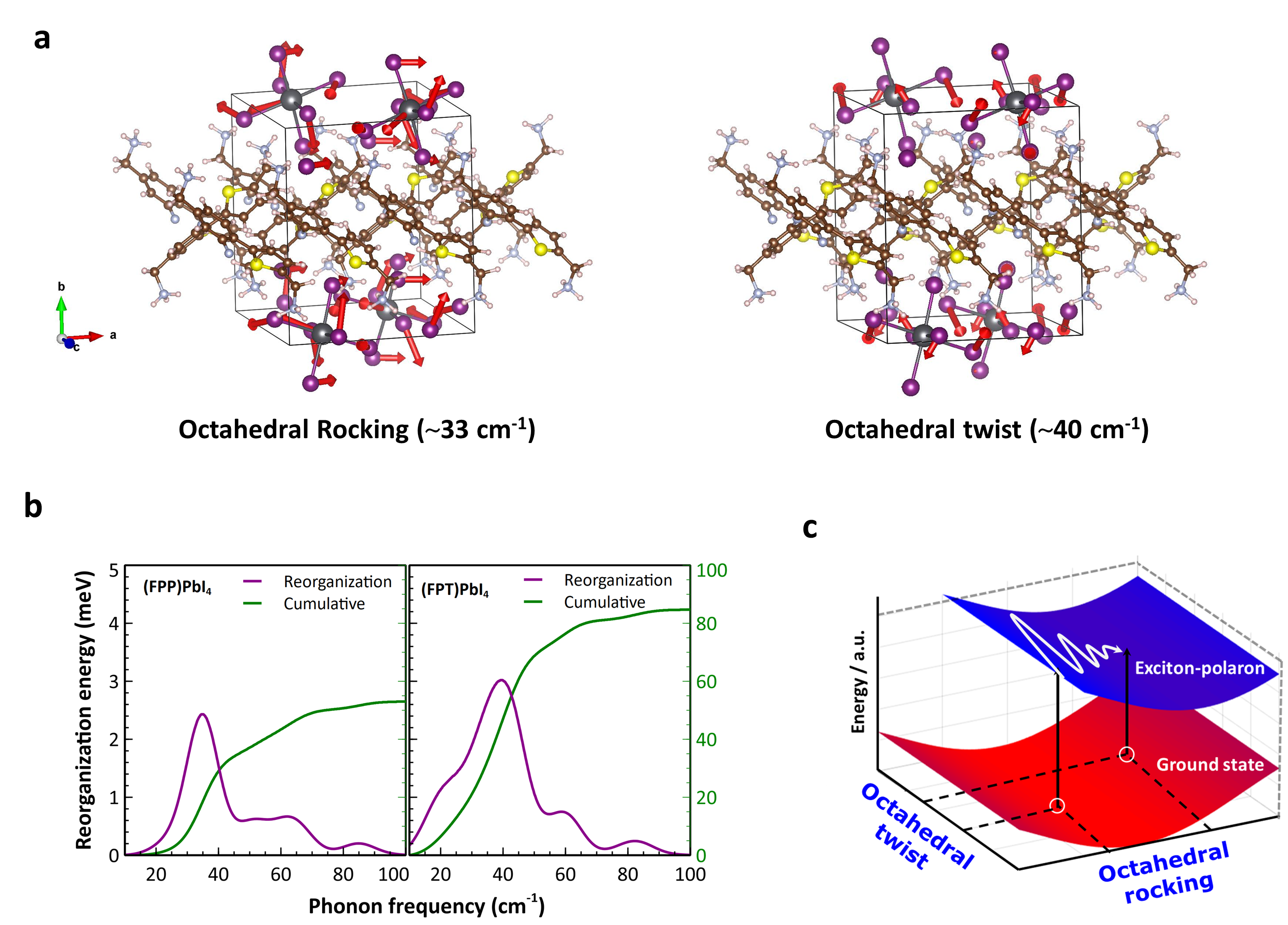}
\centering
 \caption {(a) Phonon modes that contribute significantly in exciton-polaron formation in  (FPT)\ce{PbI2} (b) Calculated reorganization energy distributions as a function of phonon frequency, indicating that excitation is strongly dressed by low-frequency phonon motions in both (FPP)\ce{PbI2} and (FPT)\ce{PbI2} (c) Schematic of the exciton-polaron formation along two phonon coordinates.}
  \label{theory}
\end{figure*}

\noindent {\large{\textbf{Estimation of exciton-phonon coupling and active participation of phonons}}}

The coherent phonon oscillation observed around 45 cm$^{-1}$ modulates the excitonic transition energy, indicating the polaronic nature of the exciton where it is dressed by the structural distortion along these phonon coordinates.
Therefore, these low-frequency modes are likely to participate in exciton-polaron formation in these materials.
In order to model the influence of low-frequency phonon motions 
we performed electronic structure calculations based on density functional theory (DFT).
We approximate exciton-phonon interactions using the special displacement method (SDM).~\cite{ZG_2016,ZG_2020}
Here, electron-phonon interactions are included by effecting special displacements of the atoms from their equilibrium sites such that the displacements yield exactly the Williams-Lax dielectric function in the thermodynamic limit.~\cite{williams,lax} 
The SDM is a quadrature scheme for efficiently performing high dimensional integrals over atomic displacements and sampling a thermal distribution assuming harmonic interactions.
Band structure calculations are performed at the static-equilibrium positions (no zero-point or thermal motion) and at displaced positions along each of the normal modes where the magnitude of the displacement is proportional to the width of the quantum harmonic thermal distribution for each mode.
Integrals over this distribution are thus approximated by a two-point quadrature.
The magnitude of the computed band structure gap shift along each normal mode gives a measure of the coupling of the different frequency normal modes to the electrons (VB) and holes (CB) that comprise the exciton.~\cite{Huang_2019}
Within the harmonic approximation, these computed gap shifts thus give a measure of the magnitude of the exciton-phonon coupling. An analysis of the relationship between the computed gap-shifts and the standard molecular definition of the reorganization energies for the different phonon modes is presented in section S7 of the supporting information.
The computational details{, and computed gap shift results} are provided in section S5 of the supplementary materials.

Figure {\ref{theory}}a shows the motion of the inorganic octahedral twist ($\sim$40 cm$^{-1}$) and rocking ($\sim$33 cm$^{-1}$) modes that contribute significantly to modulating the excitonic transition energy. 
We note that the phonon modes between 20-80 cm{$^{-1}$} strongly renormalize the gap in both (FPT){\ce{PbI4}} and (FPP){\ce{PbI4}} (Figure {\ref{theory}}b).
The calculated gap shifts are subsequently used to estimate the reorganization energies associated with the phonon modes during structural distortion (Figure {\ref{theory}}b).
The reorganization energies for the various low frequency phonon modes are on the order of $\sim$5-20 meV in both the samples (see Figure S13, the data in Figure {\ref{theory}}b have been Gaussian broadened in frequency (5 cm$^{-1}$) to connect with experiments). 
The cumulative reorganization energy of the computed low-frequency modes in (FPT)\ce{PbI4} is a factor of $\sim$1.7 larger compared to that in (FPP)\ce{PbI4}, and the calculated phonon frequencies, their effective reorganization energies, and the trends with organic ligand variation are in good qualitative agreement with the experimental results showing a broad phonon band centered at 45 cm$^{-1}$, ranging between 20-70 cm$^{-1}$ in case of (FPP){\ce{PbI4}}, while for (FPT){\ce{PbI4}} the band is slightly blue shifted to 50 cm$^{-1}$, ranging between 20-80 cm$^{-1}$.
These findings confirm that the electronic excitation is strongly coupled to the structural distortion along these phonon coordinates and establishes the polaronic nature of the exciton.
We note that the predicted DFT Kohn-Sham gap increases by only 0.01 eV going from FPP and FPT, respectively, and does not reproduce the larger shift (0.15 eV) observed in the experiment.
This may be due to errors in our estimate of the FPT structure obtained by ligand substitution and local geometry relaxation, as the detailed structure of this material is not yet known experimentally.
Nevertheless, we determine the curvature of the potential energy surface (phonon frequencies) and electron-phonon couplings that agree well with the experiment. 
A schematic of the exciton-polaron formation along these principle coordinates is shown in Figure {\ref{theory}}c. 

 In order to estimate the exciton-phonon coupling using our experimental data we adopted a widely used classical displaced harmonic oscillator model previously employed in quantum dots and perovskites.\cite{sagar2008size,Fu2021,ghosh2017free}
 According to this model, the mode-specific exciton-phonon interaction can be described using three related parameters 1) $d$ : dimensionless  lattice  displacement  of  the  normal  coordinate, 2) S : Huang-Rhys factor (S=$\frac{d^2}{2}$) for the mode, and 3) $\lambda = {\mathrm S} \hbar\omega_{ph}$ : the reorganization energy of the phonon mode with frequency, $\omega_{ph}$.
 In our experiments, the amplitude of the oscillation in the probe absorbance after excitation, A$_{osc}$, is related to the reorganization energy through A$_{osc}$ = $\frac{d OD}{ dE}\times \lambda$, and both A$_{osc}$ and optical density (OD) are consistently normalized. Estimating $\frac{d OD}{ dE}$ from the ground state absorption spectrum at various frequencies in the region around the peak oscillation amplitude we obtain estimates of the reorganization energies from the experiment and these model exciton-phonon coupling parameters are summarized in Table \ref{exciton-phonon}.

\begin{table*}
	\centering 
	\begin{tabular}{c|c|c|c|c}
		Material &$\omega_{ph} (cm^{-1})$& $\lambda (meV)$ & $S$ & $d$\\ \hline \hline
		(FPP)PbI$_4$ & 42.2 $\pm$ 0.7 & 3.48 $\pm$ 0.05& 0.67 $\pm$ 0.09 &1.16 $\pm$ 0.07\\ \hline
		
		(FPT)PbI$_4$ & 48.4 $\pm$ 1.0 & 5.58 $\pm$ 0.08& 0.93 $\pm$ 0.13 &1.36 $\pm$ 0.09\\
	\end{tabular}
	\caption{Exciton–phonon   coupling   parameters   of (FPT){\ce{PbI4}} and (FPP){\ce{PbI4}}}
	\label{exciton-phonon}
\end{table*}

These estimated Huang-Rhys factors for this class of materials are an order of magnitude larger than those reported for other 3D perovskites (S$<$0.05) and higher than the reported Huang-Rhys factor for RP-type perovskites (S= 0.11-0.30).\cite{Fu2021,ghosh2017free}
Due to the presence of quantum-well structure and dielectric confinement, the exciton-phonon coupling is generally much stronger in the case of 2D perovskites compared to 3D structures.
The estimated reorganization energies in both samples are on the order of a few meV, consistent with our electronic structural calculations.
Intriguingly, the reorganization energy and Huang-Rhys factor in the case of (FPT)PbI$_4$ are higher than that of (FPP)PbI$_4$ by a factor of 1.6 and 1.4, respectively.
Our electronic structural calculation support this observation where we find that the cumulative reorganization energy in the case of (FPT){\ce{PbI4}} is about 1.6 times higher compared to (FPP){\ce{PbI4}}. 
These results indicate the presence of higher exciton-phonon coupling and larger polaronic distortion in (FPT){\ce{PbI4}} compared to (FPP){\ce{PbI4}}. 
The observation of increased lattice relaxation during exciton-polaron formation in the case of a distorted inorganic octahedral structure (e.g see comparison of (FPT){\ce{PbI4}} and (FPP){\ce{PbI4}} optimized geometries in the SI) is in agreement with previous findings where polaronic stabilization increases with the distortion in the equilibrium lattice structure in 2D perovskites.\cite{straus2022large,zhang2023ultrafast}
This suggests that the exciton-phonon coupling and polaronic distortion in this DJ-type 2D perovskites are synthetically tunable based on the choice of the organic ligand. 
We note that the observation of the coherent phonons at a lower temperature will likely reveal a more highly resolved multi-modal energetic landscape of exciton-polaron formation.
Such analysis including lattice anharmonicity of these materials is beyond the scope of the present study and will be a subject of future investigation. 


\newpage

\noindent {\large{\textbf{Exciton-phonon interaction controls hot carrier relaxation}}}

\begin{figure*}[ht]
\includegraphics[width=6in]
{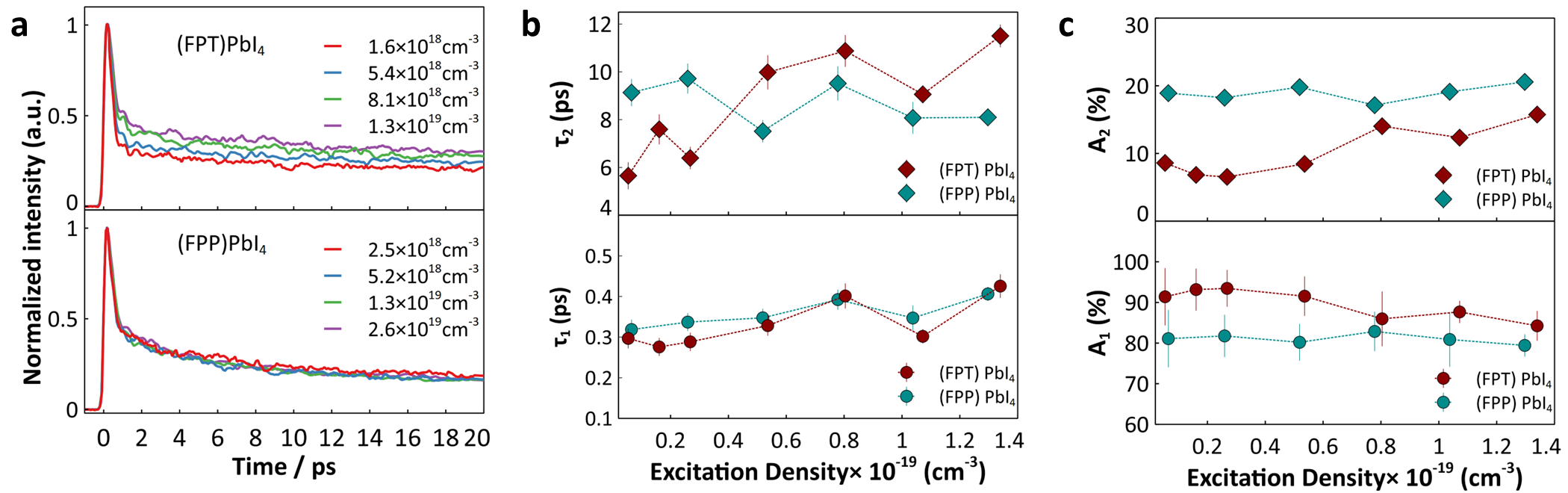}
\centering
 \caption {(a) Comparison of hot carrier cooling between (FPP)\ce{PbI4} and (FPT)\ce{PbI4} at various excitation densities, (b) Time constants ($\tau_1$ and $\tau_2$), and  (c) amplitude ($A_1$ and $A_2$)for the bi-exponential fit to the decay profiles for both (FPP)\ce{PbI4} and (FPT)\ce{PbI4}}
  \label{HCR_Comparison}
\end{figure*}

Figure \ref{HCR_Comparison}a compares the decay of the populations of hot carriers for both (FPP)PbI$_{4}$ and (FPT)PbI$_{4}$ as a function of excitation density. 
These decay profiles are obtained for the below band-gap photo-induced absorption features at 2.39 eV and 2.52 eV for (FPP)PbI$_{4}$ and (FPT)PbI$_{4}$, respectively, and are associated with hot carrier relaxation.  
The cooling of the hot carriers shows bi-exponential behavior and their fitted time constants ($\tau_1$ and $\tau_2$) and amplitudes ($A_1$ and $A_2$) are plotted in figures \ref{HCR_Comparison}b and \ref{HCR_Comparison}c, respectively.
The very fast decay component occurs on a sub-picosecond time scale with a time constant of 300-400 fs in both the materials and is independent of the excitation density (figure \ref{HCR_Comparison}b, lower panel). 
We observe a negligible change in the slower component ($\tau_2$) of the hot carrier decay as a function of excitation density in the case of (FPP)PbI$_{4}$ with an average time constant of 8.5ps and amplitude of 19\% (upper panels figures \ref{HCR_Comparison}(b,c)).
In contrast, the time constant (from 6ps to 11.5 ps) and amplitude (from 8.6\% to 15.8\%) for the long-lived component ($\tau_2$, and $A_2$), increase in the case of (FPT)PbI$_{4}$  as we increase  the excitation density from 0.2 to 1.3 $\times 10^{19}$ cm$^{-3}$ (Figure \ref{HCR_Comparison}b and c, upper panels).

 The retardation of carrier cooling at higher excitation density in the case of (FPT)PbI$_{4}$ is a signature of the hot phonon bottleneck effect.
 The slowing down of net carrier thermalization as a function of excitation density has been observed before in other 2D and 3D perovskite systems.\cite{price2015hot, yang2016observation,nie2020harnessing}
As observed in the experimental results summarized in Table \ref{exciton-phonon}, and verified in our calculations, the Huang-Rhys factor for the strongly coupled phonon modes in the  (FPT)PbI$_{4}$ system is about 40\% higher than that of (FPP)PbI$_{4}$.
 The Huang-Rhys factor can be thought of as the number of phonons dissipated to the strongly coupled modes as a result of the relaxation of the reorganization energy (S=$\frac{\lambda}{\hbar \omega_{ph}}$).\cite{huang1950theory,whalley2021giant}
 Therefore, a material with a higher Huang-Rhys factor can produce more phonons during hot carrier cooling at higher excitation density.
 Weak anharmonic couplings between phonon modes result in slow dissipation and slow thermalization suggesting the possibility of producing a phonon bottleneck.\cite{yang2016observation,price2015hot,yin2021manipulation,fu2017hot}
 At lower excitation density near 1.6 $\times 10^{18}$ cm$^{-3}$, the  (FPT)PbI$_{4}$ signal decays faster compared to (FPP)PbI$_{4}$ due to the presence of higher exciton-phonon coupling in (FPT)PbI$_{4}$.
 However, as we increase the excitation density, (FPT)PbI$_{4}$ produces a higher population of non-thermal phonons which are not able to decay sufficiently rapidly, producing the observed hot-phonon bottleneck within the range of the excitation densities considered here.
 Our study thus demonstrates that the first few picoseconds of decay of the hot carriers in these DJ-type perovskites can be controlled by incorporating different bi-functional ligands, which can sensitively vary the exciton-phonon coupling at a given excitation density.

\noindent {\large{\textbf{Competing pathways between exciton-polaron formation and hot-phonon bottleneck formation}}}

Several previous studies on both 2D and 3D perovskite systems report that polaron formation occurs within a few hundred femtoseconds following photoexcitation.\cite{zhu2016screening,seiler2022direct} 
In our data, we observe a very fast decay of the hot carriers on a sub-picosecond time scale.
Although our measurements do not  directly access the rate of exciton-polaron formation, this observation indicates that the ultrafast polaronic distortion and hot carrier relaxation may be coupled.
In order to understand whether the sub-picosecond carrier cooling ($\tau_1$) is connected to the exciton-polaron formation, we calculated the carrier temperature for the initial few picoseconds (see Figure S11).
Intriguingly we note that the sub-picosecond carrier cooling is modulated at the same frequency as the phonons that distort the lattice during exciton-polaron formation (see section S6 of supporting information for details).
This observation suggests that the sub-picosecond accumulation of the exciton-polaron population, and the ultrafast hot carrier cooling are coupled as reported recently in experimental studies using ultrafast electron diffraction measurements in perovskite nanocrystals.\cite{seiler2022direct}
As the carrier cooling and excitonic relaxation to form the polaron occur preferentially along the same normal modes, a high fraction of sub-picosecond carrier relaxation ensures a large population of the exciton-polaron state.
Therefore, the higher amplitude coefficient (A$_1$) of (FPT)PbI$_{4}$ compared to (FPP)PbI$_{4}$ may suggest a large population of the exciton-polaron state in (FPT)PbI$_{4}$ consistent with higher exciton-phonon coupling and larger polaronic stabilization in this material.
However, with increasing excitation density (0.1 to 1.4 $\times$10$^{19}$cm$^{-3}$) we observe a slight reduction (91.4 \% to 84.2 \%) of the A$_1$ amplitude in (FPT)PbI$_{4}$ with an increase (8.6 \% to 15.8 \%) in the amplitude (A$_2$) of the long-lived decay (hot-phonon population) component.
This observation suggests that the formation of a long-lived hot-phonon bottleneck may compete with the exciton-polaron population in (FPT)PbI$_{4}$.

 The microscopic origin of the long-lived hot carriers in perovskites is strongly debated and has been attributed to either large polaron formation\cite{zhu2016screening,niesner2016persistent,buizza2021polarons} or the hot-phonon bottleneck effect.\cite{yang2016observation,price2015hot,yin2021manipulation,fu2017hot}
 Our study demonstrates that although polaron formation occurs in both (FPT)PbI$_{4}$ and (FPP)PbI$_{4}$, the formation of the hot-phonon bottleneck is only observed appreciably in (FPT)PbI$_{4}$.
 We also demonstrate that the exciton-polaron and non-thermal phonon populations are controllable based on the exciton-phonon coupling within the range of excitation densities studied here.
 Therefore, our study reveals that the microscopic origin of the long-lived hot carriers can be assigned to either polaron formation or the hot phonon bottleneck depending on the exciton-phonon coupling and the excitation density in these two-dimensional DJ-type perovskite materials.
 We note that probing direct structural distortion during sub-picosecond exciton-polaron formation and how this process competes with the hot-phonon bottleneck effect requires time-resolved observation of both electronic and structural distortion using femtosecond X-ray/electron diffraction measurements.\cite{guzelturk2021visualization,cannelli2021quantifying,seiler2022direct,zhang2023ultrafast}
 However, our study identifies the phonon modes along which the exciton-polaron forms and demonstrates that exciton-polaron formation is coupled to sub-picosecond carrier cooling. 
 In particular, the ability to change the exciton-phonon coupling as a result of ligand modifications to control the exciton-polaron population, hot carrier cooling kinetics, and phonon bottleneck formation represents a significant step forward toward understanding the role of exciton-phonon coupling in 2D perovskites materials. 

 \noindent {\large{\textbf{Conclusion}}}

 Our results provide the first detailed insight into exciton-polaron formation in rigid and stable DJ-type 2D perovskite structures using ultrafast spectroscopy and electronic structure calculations. 
 We demonstrate that the excitons in this 2D perovskite material are strongly dressed by phonons to form exciton-polarons. 
 The polaronic nature of the exciton is revealed by the observation of molecular-like phonon wavepacket motion, band gap renormalization, and phonon displacement calculations.
 More importantly, we show that the exciton-phonon coupling strength can be controlled by ligand engineering, which can subsequently control the exciton-polaron population, hot-phonon bottleneck formation, and carrier cooling kinetics.
 Therefore, the rigid DJ-type 2D architecture is a suitable platform to understand the strongly coupled exciton-phonon physics as well as offering considerable promise for tuning the hot-carrier lifetime via suitable ligand engineering.

\newpage



\noindent{\large\textbf{Methods}}

\noindent{\textbf{Materials preparation and characterizations}}

\noindent The organic ligands are synthesized according to the previous report. Perovskite solutions are prepared by dissolving lead iodide and the corresponding ligand in the mixed solvent of DMF and DMSO at a volume ratio of 9:1. The thin films are fabricated by spin-coating from their corresponding solution in the ambient atmosphere. Further thermal annealing at 100$^{\circ}$C for 10 minutes is needed to give the desired 2D perovskite phase. Thin films for photodynamics study are spin-coated on glass slides, meanwhile, thin films for X-ray diffraction (XRD) and scanning electron microscopy (SEM) study are spin-coated on silicon wafer. Both substrates are cleaned by ultrasonication in deionized water, acetone, and isopropanol, sequentially, and 10 minutes for each step. Then dried in the oven. Prior to use, the substrates are treated with plasma cleaner for 20 minutes. The crystallinity and phase purity of the thin films are investigated by measuring their XRD patterns (See SI section 1). Compared with thin films spin-coated from DMF solution, this batch of thin films shows improved crystallinity potentially due to the longer crystallization time induced by the addition of higher boiling point DMSO. Film morphology was explored using scanning electron microscopy (SEM) (See SI section 1). Corresponding to the enhanced crystallinity, crystalline domains are larger in size than in thin films spin-coated from pure DMF.

\noindent{\textbf{Ultrafast Measurements}}

\noindent Ultrafast measurements were performed using a femtosecond transient absorption setup. 
The output (800 nm/1.55 eV at 1 kHz with 45-fs pulse duration, average power $\sim$4 W) of a commercial Ti: sapphire laser (Coherent, Libra) was split using a beam splitter to generate the pump and probe pulses.
The pump beam was produced by driving an optical parametric amplifier (OPerA Solo) to convert the fundamental 800 nm radiation to the desired wavelength of interest for photoexcitation.
The data shown in Figure 1 and Figure 4 of the main text were obtained using a photoexcitation wavelength of 410 nm with a pulse duration of about $\sim$100 fs.
This pulse excites electrons along with the creation of vibrational (phonon) wavepackets. 
 A neutral density filter was used to control the pump power for the fluence dependence study.
 Another part of 800 nm fundamental was focused on a \ce{CaF2} crystal to produce a broadband probe beam between 320nm (3.87 eV) to 700nm (1.77 eV).
 This region is suitable to probe the modulation of excitonic resonance for both (FPP)\ce{PbI4} ($\sim$2.47 eV) and (FPT)\ce{PbI4} ($\sim$2.63 eV).
 The pump and probe pulses were focused on the sample at the near-normal incidence and the measurements were performed in a transmission geometry using a commercial pump-probe setup (Ultrafast System Helios).
 The transmitted probe beam was focused into a fiber optic using a lens that is connected to a visible sensor detector.
 The time delay between the pump and probe beam was varied using a retroreflector mounted on a delay stage.

 \noindent{\textbf{Computational Details}}

 \noindent Density functional theory (DFT) calculations were performed on (FPP)\ce{PbI4}  and (FPT)\ce{PbI4}  using the Quantum Espresso package within the Perdew-Burke-Ernzerhof (PBE)  exchange-correlation functional. To account for intermolecular van der Waals interactions, the Tkatchenko-Scheffler (TS) dispersion correction was applied. The core and nuclear interactions were described by the Optimized Norm-Conserving Vanderbilt (ONCV) pseudopotentials. For the two crystals, geometry optimization had a maximum force of $1\times10^{-4}$ Ry/Bohr with a force and total energy convergence thresholds of $1\times10^{-7}$ Ry/Bohr and of $1\times10^{-9}$ Ry, respectively.  The lattice parameters post geometry optimization were predicted to be $a= \mathrm{11.95 \AA}$, $b=\mathrm{14.28 \AA}$ and $c=\mathrm{12.80 \AA}$ for (FPP)\ce{PbI4}, and $a= \mathrm{12.11 \AA}$, $b=\mathrm{13.89 \AA}$ and $c=\mathrm{12.82 \AA}$ for (FPT)\ce{PbI4}.

 \noindent To capture electron-phonon interactions at finite temperature and the quantum zero-point renormalization to the band gap energy, we applied the special displacement method developed by Zacharias and Giustino (ZG). In this method, the ZG configuration captures electron-phonon interactions by applying a special set of displacements that reproduce thermodynamic disorder of the nuclei positions. The ZG configurations of individual phonon modes were computed using the ZG.x code as part of the EPW module within Quantum Espresso~\cite{Hyungjun2023}. The single-mode ZG configurations were generated at T=300K using the code version 7.1, where the imaginary frequencies are “frozen” and do not contribute to the bandgap renormalization. More details are provided in the SI.

\noindent{\textbf{ Data availability}}

\noindent The authors declare that all data supporting the findings of this study
are available within the paper and Supplementary Information files.
Additional images are available from the corresponding authors
upon request.

\noindent{\textbf{Acknowledgement}}

\noindent Financial support was provided by the Division of Chemical Sciences, Geosciences and Biosciences, Office of Basic Energy Sciences, of the US Department of Energy through grant no. DE-SC0015429. Additionally, this work was supported by the US Department of Energy, Office of Basic Energy Sciences under contract DE-SC0020437. DFC acknowledges support from the National Science Foundation under grant CHE-1955407. FA acknowledges support from the Department of Energy, Office of Basic Energy Sciences under contract DE-SC0018080. SS acknowledges support from the Department of Energy, Office of Basic Energy Sciences under contract DE-SC0023402. R.Z. and D.S.S. acknowledge support from the NSERC of Canada and the E.W.R. Steacie Memorial Fellowship. 

\noindent{\textbf{Author contributions}}

\noindent S.B., G.D.S, and D.S.S initiated the project. S.B. designed the study, performed ultrafast experiments, and analyzed the experimental data. F.A., M.Z., S.S., and D.F.C. designed the computational study, ran calculations, and analyzed the theoretical results. R.Z. and D.S.S designed and fabricated the materials. S.B., G.D.S, and D.F.C extensively discussed the data. All authors contributed to writing the manuscript.

\noindent{\textbf{Competing interests}}

\noindent The authors declare no competing interests.

\noindent{\textbf{Additional information}}


\noindent \textbf{Correspondence and requests for materials} should be addressed to Gregory D. Scholes.

\bibliography{achemso-demo}

\end{document}